\documentclass[runningheads]{llncs}

\usepackage{eccv}

\usepackage{eccvabbrv}

\usepackage{graphicx}
\usepackage{booktabs}

\usepackage{amsmath}
\usepackage[utf8]{inputenc}
\usepackage{geometry}
\usepackage{array}
\usepackage{booktabs}
\usepackage{multirow}

\usepackage{caption}
\usepackage{subcaption}

\usepackage[accsupp]{axessibility}  

\usepackage{hyperref}

\usepackage{orcidlink}

\begin{document}

\title{Going Beyond U-Net: Assessing Vision
Transformers for Semantic Segmentation
in Microscopy Image Analysis} 

\titlerunning{Assessing Vision
Transformers for Semantic Segmentation}

\author{Illia Tsiporenko\inst{1}\orcidlink{0009-0009-0404-0679}\and
Pavel Chizhov\inst{1,2}\orcidlink{0009-0001-7329-6899}
\and
Dmytro Fishman\inst{1, 3}\orcidlink{0000-0002-4644-8893}}

\authorrunning{Tsiporenko et al.}


\institute{
    \textsuperscript{1}Institute of Computer Science, University of Tartu\\
    \email{\{illia.tsiporenko, pavel.chizhov, dmytro.fishman\}@ut.ee}\\
    \url{https://ut.ee}\\
    \textsuperscript{2}Center for Artificial Intelligence, Technical University of Applied Sciences Würzburg-Schweinfurt\\
    \email{pavel.chizhov@thws.de}\\
    \url{https://thws.de} \\
    \textsuperscript{3}STACC, Estonia, Tartu \\
    \url{https://stacc.ee/} \\
}
\maketitle

\begin{abstract}
Segmentation is a crucial step in microscopy image analysis. Numerous approaches have been developed over the past years, ranging from classical segmentation algorithms to advanced deep learning models. While U-Net remains one of the most popular and well-established models for biomedical segmentation tasks, recently developed transformer-based models promise to enhance the segmentation process of microscopy images. In this work, we assess the efficacy of transformers, including UNETR, the Segment Anything Model, and Swin-UPerNet, and compare them with the well-established U-Net model across various image modalities such as electron microscopy, brightfield, histopathology, and phase-contrast. Our evaluation identifies several limitations in the original Swin Transformer model, which we address through architectural modifications to optimise its performance. The results demonstrate that these modifications improve segmentation performance compared to the classical U-Net model and the unmodified Swin-UPerNet. This comparative analysis highlights the promise of transformer models for advancing biomedical image segmentation. It demonstrates that their efficiency and applicability can be improved with careful modifications, facilitating their future use in microscopy image analysis tools.
\end{abstract}

\section{Introduction}
Identifying objects in microscopy images is a crucial first step in successful image analysis~\cite{microscopy_intro}. Precise segmentation of various cellular structures, including nuclei, enables the extraction and analysis of vital morphological features. However, achieving accurate and efficient segmentation remains challenging due to the complex and heterogeneous nature of microscopy data.

Deep learning algorithms are powerful tools for segmentation tasks, given their ability to generalise and understand underlying image structures. For a long time, the traditional Convolutional Neural Network (CNN) U-Net~\cite{Unet} has been one of the most popular and well-established models in this field, demonstrating notable results in various microscopy image segmentation tasks. However, many new deep learning models have been developed over the past few years, with transformers among the most promising. Transformers use the attention mechanism~\cite{attention_is_all_you_need} at their core, allowing them to capture complex image structures, provide an unlimited receptive field, and incorporate more local context than traditional CNNs. These features are particularly advantageous for enhancing the segmentation process of microscopy images, where capturing local context is essential for improving the finer details in segmentation masks.

This paper provides an assessment of segmentation models, which utilise some of the most popular vision transformers as image encoders --- Vision Transformer~\cite{ViT} (ViT), present in the UNETR~\cite{UNETR} model, and Swin Transformer~\cite{SWIN}, present in the Swin-UPerNet~\cite{SWIN} model. Additionally, we assess the novel foundational Segment Anything Model (SAM), which uses user-defined prompts to enhance the segmentation process. We compare these models to the robust and lightweight U-Net model, which serves as our baseline.

Swin Transformer in combination with UPerNet-based decoder demonstrated promising performance in semantic image segmenation~\cite{SWIN}. However, the model's reliance on processing image patches of size 4 inevitably leads to the loss of fine-grained details. This loss of low-level details, coupled with the use of bilinear interpolation in the decoder, may compromise the overall performance of the model by reducing the precision of the segmentation boundaries and affecting the accuracy of object delineation. In our work, we aim to address this issue by designing encoder and decoder enhancements to introduce local context and improve the flexibility of mask generation, thereby improving detail capture and segmentation accuracy.

By studying the capabilities of these transformer-based models, we aim to highlight their potential advantages and drawbacks compared to the traditional U-Net model. In our comparative analysis, we seek to demonstrate the promise of transformers in advancing microscopy image segmentation.

\section{Related Work}

While U-Net~\cite{Unet} remains one of the most popular models for segmentation tasks in the biomedical domain, recent years have seen the development of many new transformer-based models~\cite{SWIN, swinUNETR, Swin-Unet, transunet, sam, UNETR, SETR, hybrid_trans_unet,sunet,SSFormer}. These models can be roughly divided into two broad categories: transformer-CNN and hybrid models. Transformer-CNN models use transformers as the primary image encoder, while CNN layers in the decoder generate the segmentation masks. Examples include UNETR~\cite{UNETR}, Swin UNETR~\cite{swinUNETR}, Swin-UPerNet~\cite{SWIN}, and SETR~\cite{SETR}. On the other hand, hybrid models utilize both CNN and transformer layers in the encoder but retain CNN layers in the decoder. An example of this model type are TransUNet~\cite{transunet}, SU-Net~\cite{sunet}, and CS-UNet~\cite{CS-Unet}. 

Even though hybrid models are more flexible in design and allow for more architectural experiments, one major advantage of Transformer-CNN models is the use of intact transformer encoders pre-trained on large datasets like ImageNet~\cite{ImageNet}. Such models generally show superior performance over hybrid ones, as the improvement coming from transformers is mostly related to large and diverse pre-training~\cite{SWIN}. This difference renders hybrid models less preferable, thus we decided to omit them in our experiments.

There is also a separate category of models that have been recently gaining popularity~--- foundational models. These models are typically trained on massive datasets and offer zero-shot generalisation. Such a model was recently introduced for image segmentation --- Segment Anything Model~\cite{sam} (SAM). SAM uses user-defined prompts, such as bounding boxes or points, to guide and improve the segmentation process.
 
As the Swin Transformer has demonstrated superior performance in many imaging tasks, numerous new re developed using Swin as a basis~\cite{transunet, swinUNETR, Swin-Unet}. Swin-UPerNet was the first model that used Swin as the encoder in combination with the UPerNet decoder. Following its success, many other segmentation models that employ Swin as the backbone were developed. Some propose different types of decoders compared to the original Swin-UPerNet, such as Swin UNETR~\cite{swinUNETR} and SSformer~\cite{SSFormer}, while others follow the idea of hybrid models such as CS-UNet~\cite{CS-Unet}, where both the encoder and decoder are revised. However, to the best of our knowledge, there is a notable gap in research on the original Swin-UPerNet, with opportunities for enhancing this model. Thus, in this work, we explore Swin-UPerNet, identify its potential limitations, and address them through custom modifications. These modifications, while greatly improving its performance, preserve the original architecture of the model, enabling the reuse of the pre-trained weights, which improves the convergence of the loss during training and enhances the overall performance of the model.

\section{Methods}
Our work aims to comprehensively compare the well-established U-Net model and notable transformer-based models, including UNETR~\cite{UNETR}, Swin-UPerNet~\cite{SWIN}, and the Segment Anything Model~\cite{sam}, specifically within the microscopy domain. Additionally, we design custom modifications for Swin-UPerNet to enhance its performance in microscopy image segmentation tasks. In this section, we will describe the datasets, the configuration of the models, and the approaches for training and evaluation.

\begin{table}[t]
\newcommand{\mr}[1]{\multirow{2}{*}{#1}}
  \caption{Detailed overview of datasets used in the study. Here, we detail the number of images present in each dataset, the resolution and the number of channels of each image in the dataset, the segmentation target and the modality of the images.}
    \label{tab:datasets}
  \centering
  \setlength{\tabcolsep}{0.25cm}
  \begin{tabular}{cccccc}
    \toprule
       \mr{Dataset} & Images & \mr{Resolution} & \mr{Channels} & \mr{Target} & \mr{Modality} \\
                    & (Train / Test) &         &               &             &               \\
    \midrule
      Seven    & \mr{2016 / 504} & \mr{$1080\times1080$} & \mr{1} &  \mr{Nuclei}  &  \mr{Brightfield}\\ 
    Cell Lines &           &                       &        &               &                  \\
    \midrule
     Electron  & \mr{366 / 99} & \mr{Varies} & \mr{3} & \mr{Varies} &  Electron \\
    Microscopy &          &             &        &             & Microscopy \\
    \midrule
    \mr{LIVECell} & \mr{3253 / 1986} & \mr{$768\times512$} & \mr{1} & Individual &   Phase \\
                  &           &                     &        &    Cells   &  Contrast \\
    \midrule
    MoNuSeg & 250 / 140 & $512\times512$ & 3 & Nuclei & Histopathology \\
    \bottomrule
  \end{tabular}
\end{table}

\subsection{Datasets}
\begin{figure}[t]
    \centering
    \begin{subfigure}{0.24\textwidth}
        \centering
        \includegraphics[width=\textwidth]{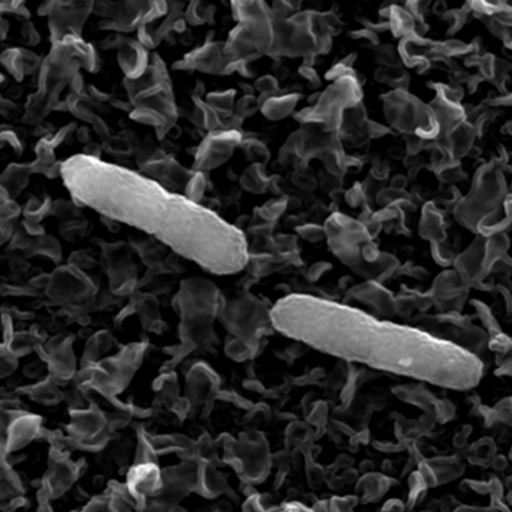}
        \caption{}
        \label{em_fig}
    \end{subfigure}
    \begin{subfigure}{0.24\textwidth}
        \centering
        \includegraphics[width=\textwidth]{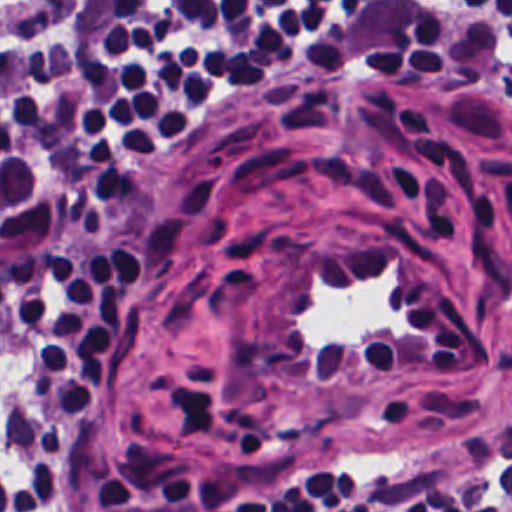}
        \caption{}
        \label{monuseg_im}
    \end{subfigure}
    \begin{subfigure}{0.24\textwidth}
        \centering
        \includegraphics[width=\textwidth]{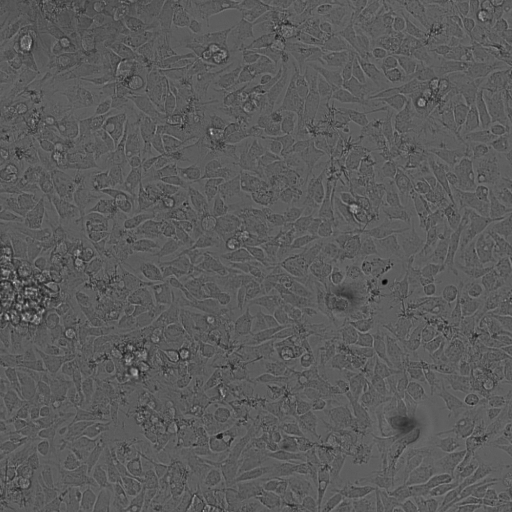}
        \caption{}
        \label{seven_img}
    \end{subfigure}
    \begin{subfigure}{0.24\textwidth}
        \centering
        \includegraphics[width=\textwidth]{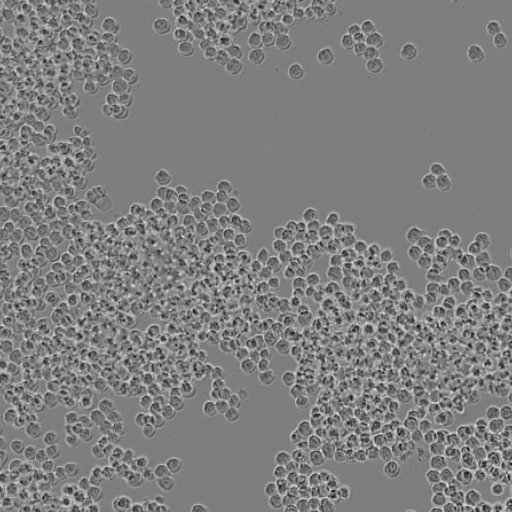}
        \caption{}
        \label{livecell_ing}
    \end{subfigure}
    \caption{We present example crops of the images from (a) Electron Microscopy dataset~\cite{em}, (b) MoNuSeg dataset~\cite{monuseg_1, monuseg_2}, (c) Seven Cell Lines dataset~\cite{microscopy_intro}, and (d) LIVECell dataset~\cite{livecell}}
    \label{datasets_fig}
\end{figure}

 To assess the performance and capabilities of those models, we chose four different datasets, each representing a distinct image modality, offering unique segmentation challenges. Table \textcolor{red}{\ref{tab:datasets}} provides a detailed overview of the chosen datasets and Figure \textcolor{red}{\ref{datasets_fig}} provides an example image from each dataset. The Electron Microscopy dataset~\cite{em} contains images of various resolutions, focusing on electron microscopy image modality. The Seven Cell Lines dataset~\cite{microscopy_intro} contains brightfield images with a resolution of $1024\times1024$, targeting nuclei of cells. The LIVECell dataset~\cite{livecell} consists of phase-contrast images of $768\times512$, mainly focusing on individual cells. MoNuSeg dataset~\cite{monuseg_1}~\cite{monuseg_2} includes whole-slide histopathology images, which we tiled into smaller images of size $512\times512$, with the main target being nuclei of tissue cells. This thorough collection of datasets allows us to fairly evaluate the capabilities of each of the segmentation models in various segmentation scenarios, ensuring an in-depth assessment.

\subsection{Segmentation Models}
 The U-Net~\cite{Unet} model serves as our practical baseline for this study as it is known for its robust and notable performance in microscopy image segmentation tasks. Its architecture was specifically designed for biomedical imaging tasks, featuring a symmetric encoder-decoder structure with skip connections.

 For transformer-based models, we chose the models that employ different types of vision transformer encoders and different approaches for segmentation to assess their advantages and limitations. Among many transformer-based image encoders, two are well-established in the field --- the ViT~\cite{ViT} and the Swin Transformer~\cite{SWIN}.
 The first transformer model we chose for assessment was UNETR~\cite{UNETR}, which was initially designed for 3D biomedical image segmentation. We adapted UNETR for 2D image segmentation to use in our experiments. UNETR utilises ViT at its core to encode the images. The decoder part is similar to the U-Net model, consisting of a series of convolutional layers and transposed convolutions to upscale the feature maps produced by the encoder. On the other hand, Swin-UPerNet~\cite{SWIN} utilises a different type of encoder --- Swin Transformer, with its unique windowed attention mechanism and patch merging operations making it possible to extract the features from the input image on different scales. UPerNet serves as the decoder of the network, consisting of a Feature Pyramid Network~\cite{FPN} (FPN), a Pyramid Pooling Module~\cite{PPM} (PPM), and the final upscaling layer --- bilinear interpolation. Lastly, the Segment Anything Model~\cite{sam} presents a unique approach to the segmentation tasks by utilising user prompts, such as points or bounding boxes, enhancing the performance of the model in complex microscopy image segmentation tasks. 

 \subsubsection{U-Net}
 We utilised the Segmentation Models Pytorch~\cite{smp} (SMP) framework to construct the U-Net model. ResNet34, pre-trained on the ImageNet dataset~\cite{ImageNet}, was used as the backbone for the network. We kept the network parameters and configuration as predefined in the framework. The depth of the encoder was set to 5 stages, where each stage generates feature maps two times smaller in spatial dimension than the previous one. 

 \subsubsection{UNETR}

We adapted the original version of UNETR~\cite{UNETR}, designed for biomedical tasks, to handle 2D microscopy images. We followed the same original architectural ideas of the model with slight adjustments --- all of Conv3D layers in the decoder part of the network were replaced with Conv2D. We utilised the base version of ViT, pre-trained on ImageNet dataset~\cite{ImageNet}, with the patch size of $16\times16$

\subsubsection{Segment Anything Model}
We utilised SAM~\cite{sam} out of the box, pre-trained on the SA-1B dataset~\cite{sam}. We assessed all of the ways to segment images with SAM: automatic segmentation, providing user-defined point or bounding box prompts. The bounding boxes represent the highest degree of user interaction with the model and, thus --- the highest degree of effort compared to the point prompting. The model expects bounding boxes as input in the [$\text{B} \times 4$] format, where $\text{B}$ represents the number of output masks. Similarly, the input format for point prompts is [$\text{B} \times \text{N} \times 2$], where B is the number of output masks, and N represents the number of points per object. On the other hand, automatic segmentation requires no interaction with the model from the user side, segmenting all potential objects and structures in the image. To assess the performance of the model, we used the OpenCV~\cite{opencv} framework to generate relevant points and bounding boxes from the ground truth masks, which were provided in the datasets. These prompts served as the input to the model alongside the corresponding image to obtain final results.

\subsubsection{Swin-UPerNet}
We used the Swin-UPerNet~\cite{SWIN} model, pre-trained on the ImageNet dataset~\cite{ImageNet}. We utilised the small (Swin-S) and base (Swin-B) versions of the Swin Transformer for the encoder. The decoder remained the same, consisting of FPN, PPM, and final linear interpolation. The default configuration of Swin-UPerNet uses a patch size of $ 4\times4$ with a window of size 7 (Table \textcolor{red}{\ref{tab:specs}} provides a detailed overview of configuration).

\begin{table}[t]
\centering
\caption{Detailed overview of parameters of the small (Swin-S) and base (Swin-B) versions of Swin-UPerNet.}
\label{tab:specs}
\setlength{\tabcolsep}{0.25cm}
\begin{tabular}{lll}
\toprule
\textbf{Parameter}   & \textbf{Swin-S} & \textbf{Swin-B}\\
\midrule
Patch size           & $4 \times 4$ & $4 \times 4$ \\
Embedding dimension  & 96           & 128 \\
Window size          & 7            & 12 \\
Depth of transformer & 2, 2, 18, 2  & 2, 2, 18, 2 \\
Heads in each stage  & 3, 6, 12, 24 & 4, 8, 16, 32 \\
Hidden size in MLP   & 768          & 1024 \\
\bottomrule
\end{tabular}
\end{table}

\subsection{Swin-UPerNet Modifications}

\begin{figure}[!htbp]
    \centering
    \includegraphics[width=0.8\linewidth]{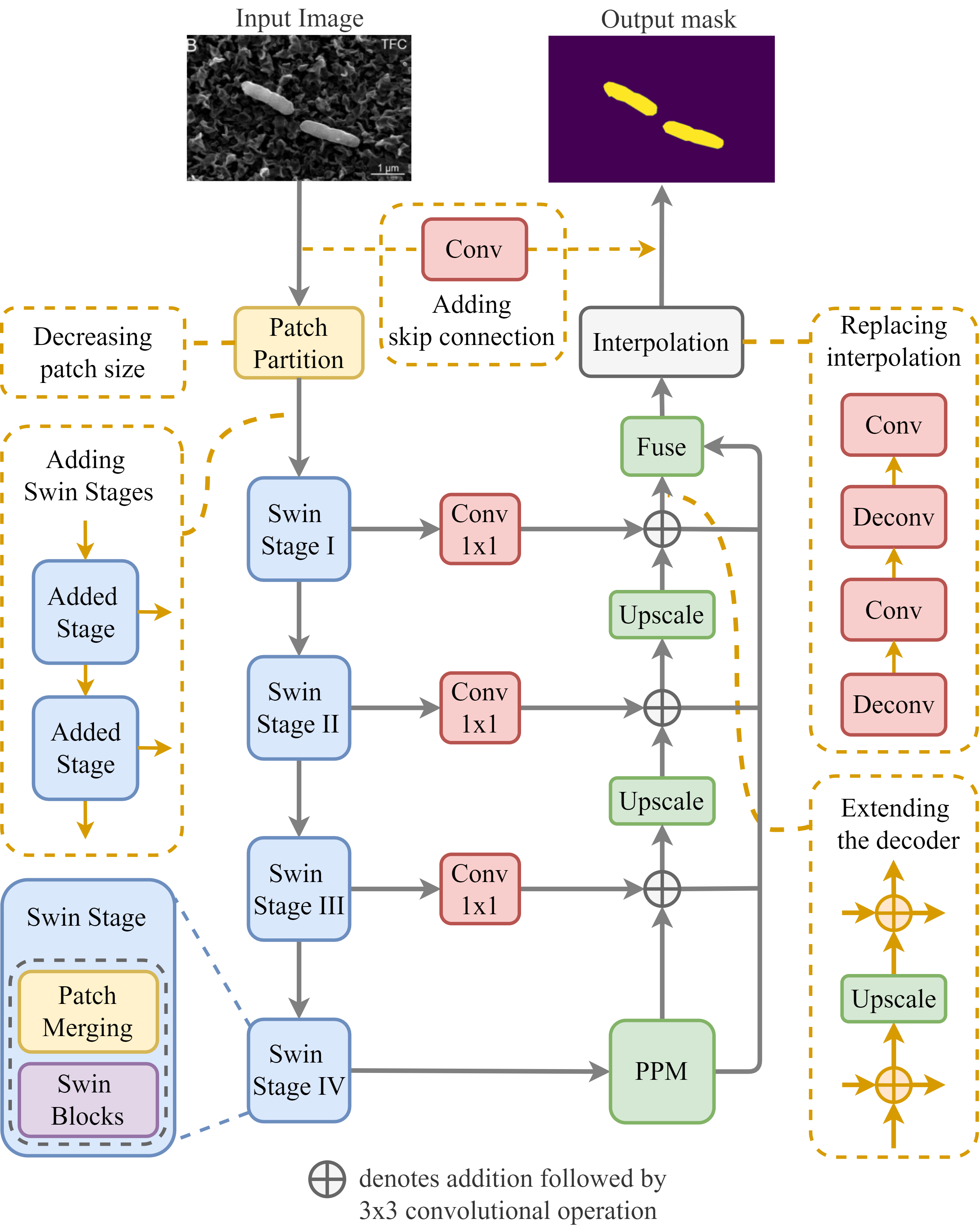}
    \caption{Representation of Swin-UPerNet architecture, which consists of Swin Transformer (blue blocks) and the UPerNet decoder (green blocks). Orange dotted rectangles provide an overview of our proposed modifications to the architecture of the model. Conv denotes a convolutional block, which is made of a convolutional layer, batch normalisation, and ReLU activation. Deconv denotes transposed convolutional operation. The circle with a line denotes an addition operation, followed by a convolutional operation with kernel size $3\times3$.}
    \label{fig:architecture}
\end{figure}

While exploring the Swin-UPerNet, we identified several issues in the network. As the original model uses a patch size of $4\times4$, the input size reduces by 4 times after the patch partitioning operation. This causes the misalignment in the decoder part of the network. In order to align the dimension of the final segmentation mask with the input image, bilinear interpolation is used in the original implementation of the model. While this approach provides a clear and lightweight solution, it has some drawbacks. Bilinear interpolation does not have learnable parameters and can introduce artefacts in the segmentation mask, potentially decreasing the performance of the model. To address this issue, we propose an architectural improvement by replacing the bilinear interpolation with a series of convolutional and transposed convolutional layers, introducing more learnable parameters and enhancing the quality of the segmentation mask.

As the segmentation tasks can be challenging when dealing with microscopy images, and the size of the objects may greatly vary, ranging from tiny nuclei to whole cells, it is necessary to induce more local information. We proposed different ways to achieve it and enhance the performance of Swin-UPerNet, specifically in microscopy image segmentation tasks:
\begin{enumerate}
    \item Decreasing the patch size to induce more local context.
    \item Adding a skip connection with a convolutional block from the input image to the decoder part of the network.
    \item Adding additional Swin Transformer stages into the backbone. 
\end{enumerate}

To address these issues, we designed several architectural and configurational improvements. 
Figure \textcolor{red}{\ref{fig:architecture}} illustrates the main ideas of our designed modifications, and Table \textcolor{red}{\ref{swin_mods}} provides an overview of modification present in different types of proposed architectures.

These modifications aim to induce more local context in the model and potentially increase its performance. We present a detailed overview and explanation of each designed modfication below. 
\subsubsection{Swin-S-PS2} We changed the patch size from $4\times4$ to $2\times2$, increasing the ability of the model to capture finer details. We kept all other parameters the same.

\subsubsection{Swin-S-Conv} We replaced the bilinear interpolation, which has no learnable parameters and sometimes produces artefacts in the final segmentation mask, with a series of convolutional blocks alongside transposed convolutions. Each convolutional block consists of a Conv2D layer with a kernel size of $3\times3$ and padding of 1, Batch normalisation, and ReLU activation function. We further enhanced the model by adding a skip connection with a convolutional block and merging its feature maps with those generated by the decoder. This set of modifications aims to enhance the quality of the final segmentation mask and increase the performance of the model.

\subsubsection{Swin-S-TB}
We decreased the patch size of $2\times2$ and kept the convolutional and transposed convolutions in the decoder part. Additionally, we integrated one more stage in the backbone of the network, consisting of two consecutive Swin Transformer blocks. All of those changes aim to increase the ability of the model to process complex features and induce more of the local context.

\subsubsection{Swin-S-TB-Skip} We kept the previous ideas of Swin-S-TB. Additionally, we added a skip connection in order to introduce more low-level information and see how much it contributes to the quality of the final segmentation mask.

\subsubsection{Swin-S-Pyramid}\label{pyramid} We decided to decrease the patch size even more --- to $1\times1$. We extended the backbone of the network, adding two additional stages. We changed the embedding dimension to 24 in order to align with the desired input of the rest of the backbone, keeping the pre-trained weights. The architecture of the decoder was adjusted so that the output of two additional stages in the backbone aligns with the FPN, yielding the final segmentation mask with the same dimensions as the input image. With these changes, we do not need to have any additional convolutional layers or interpolation in the decoder.

\begin{table}[t]
    \caption{Detailed overview of Swin-UPerNet modifications. Each row represents the Swin-UPerNet modification. The checkmarks show modifications present in the architecture. \textbf{Deconv2D} denotes a series of convolutional blocks and transposed convolutional layers. \textbf{Skip} denotes the presence of a skip connection from the input image to the decoder part of the network. \textbf{Extra Stage} denotes the additional Swin stages in the encoder of the network. \textbf{Pyramid} denotes the modification with an extended encoder and decoder.}
    \label{swin_mods}
    \centering
    \setlength{\tabcolsep}{0.25cm}
    \begin{tabular}{lccccc}
    \toprule
         \textbf{Models}        & \multicolumn{4}{c}{\textbf{Modifications}} \\
         \addlinespace[4pt]
                      & Patch & \multirow{2}{*}{Deconv2D} & Skip           & Extra & \\
                      & Size  &                           & Connection     & Stage & \\
         \midrule         
         Swin-UPerNet & $4\times4$ & --- & --- & --- \\
         Swin-S-PS2     & $2\times2$ & --- & --- & --- \\
         Swin-S-Conv    & $4\times4$ & \checkmark & \checkmark & --- \\
         Swin-S-Pyramid & $1\times1$ & --- & \checkmark & \checkmark  \\
         Swin-S-TB      & $2\times2$ & \checkmark & --- & \checkmark  \\
         Swin-S-TB-Skip & $2\times2$ & \checkmark & \checkmark & \checkmark \\
         \bottomrule
    \end{tabular}
\end{table}

\subsection{Training Pipeline}
\label{training_pipeline}

We designed our custom training pipeline to effectively train and switch between different deep learning models and their modifications. We utilised Pytorch~\cite{pytroch}, Weights and Biases~\cite{wandb}, and Hydra~\cite{Hydra} frameworks for flexible training, configuration management, tracking and logging the experiments.

\subsubsection{Data Preprocessing}

Input images are normalised and transformed using the Albumentations~\cite{albumnetations} framework. We apply horizontal and vertical flips and random rotation to the input. This choice enhances the ability of the model to understand the structure of the images across different orientations and scales of the input images, which often can be the case in microscopy images.

Additionally, we use the random cropping of size $224\times224$ during the training process. We found this crop size beneficial as it allows us to avoid unnecessary padding during training Swin-UPerNet and other transformer-based models. If the height and the width of the input image were not multiple of the product of window size and scaling factors across the layers of the network, the additional padding is applied to process the image. This padding can lead to artefacts in the final segmentation mask and potentially affect the performance of the model.

\subsubsection{Training}

Each model was trained for 150 epochs, which we found optimal for convergence. The batch size of 16 images was used --- the maximum that could fit into our GPU memory. We sampled 500 images from the dataset during training each epoch to provide diverse examples and enhance the robustness of the model. We chose the combination of Dice~\cite{DiceLoss} and Focal~\cite{FocalLoss} losses for training with weight coefficients set to 0.9 and 0.1, serving as the standard ratio. We compute the loss as follows:

 \begin{equation}
     \mathcal{L}_{total} = \alpha\times\mathcal{L}_{dice}(Y, \hat{Y}) + \beta\times\mathcal{L}_{focal}(Y, \hat{Y})
 \end{equation}
Here $Y$ is the ground truth mask, $\hat{Y}$ is the predicted mask, and $\alpha$ and $\beta$ are the weight coefficients.

\subsubsection{Evaluation}

For the evaluation, we used the F1 score and IoU score as our primary metrics to thoroughly assess the performance of all models. The evaluation itself was done on separate test sets of full-size images. For the UNETR and U-Net models, we applied a custom tiling algorithm --- the input image was divided into tiles, and the model predicted the segmentation mask for each of the tiles. Those segmentation masks of tiles were merged back to obtain the final full-size segmentation mask. 

\subsection{Computational Resourses}

We trained all of the models on the High-Performance Computing Cluster of the University of Tartu~\cite{hpc}, which has Nvidia Tesla V100 GPUs with 32 gigabytes of VRAM and Nvidia Tesla A100 GPUs with 40 and 80 gigabytes of VRAM running CUDA 12.3 with Driver version 545.23.08.

\begin{figure}[!htbp]
    \centering 
    \includegraphics[width=0.66\linewidth]{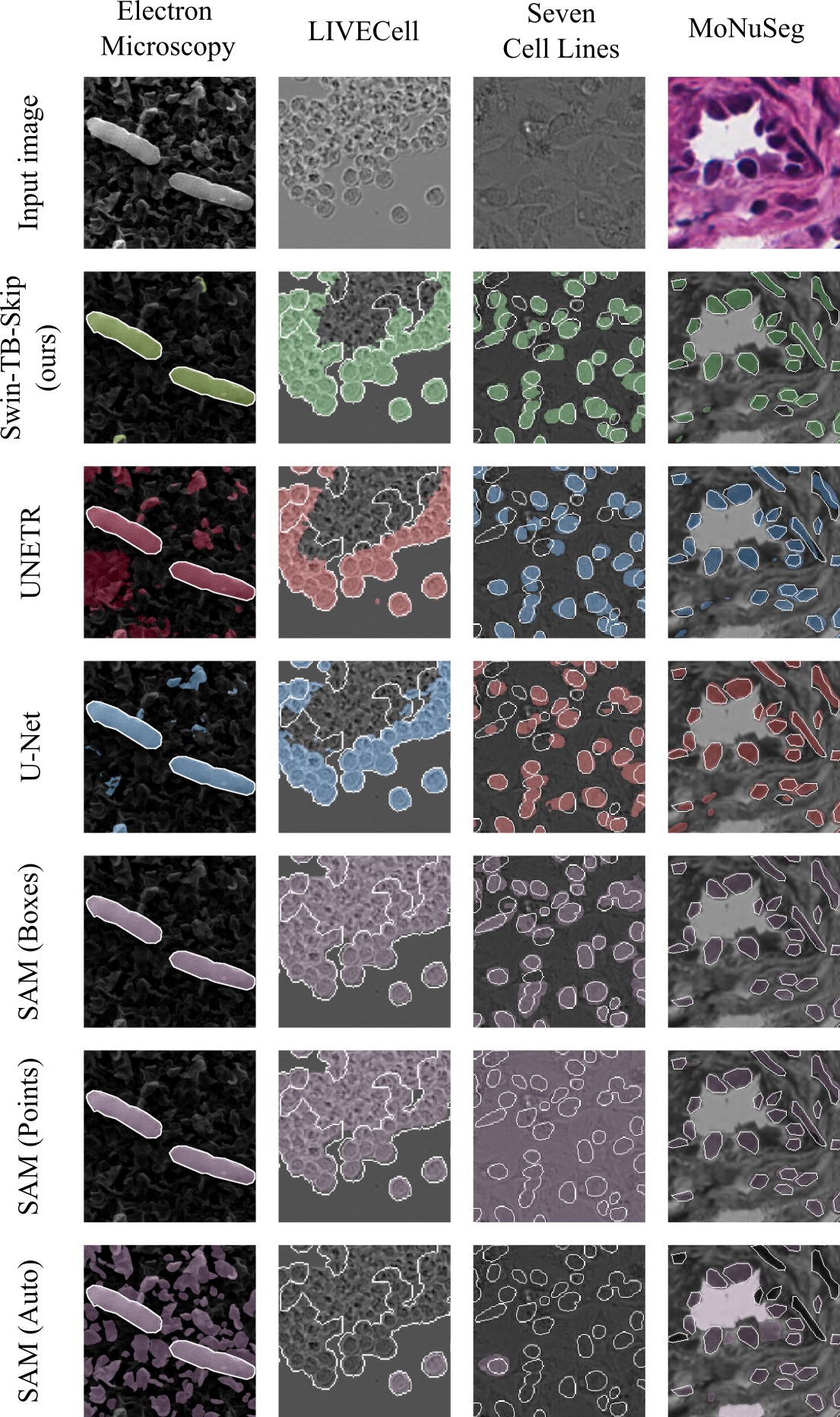}    \caption{Predicted segmentation masks of Swin-S-TB-Skip, UNETR, U-Net, and Segment Anything Model (utilising bounding box and point prompts and enabling automatic segmentation). The white contour represents the outline of the ground truth mask. The colour overlay represents the predicted segmentation mask of the model: green colour for Swin-S-TB-Skip, red colour for UNETR, blue colour for U-Net, and purple colour for SAM. We made the image from MoNuSeg dataset grayscale for the purpose of better visibility of predicted segmentation masks.}
    \label{fig:examples}
\end{figure}

\section{Results}

Here, we present the results of our experiments. Firstly, we compare the U-Net model against the chosen transformer-based models --- UNETR, Swin-S and Swin-B, and SAM. Next, we will present and detail the results of our designed modifications for Swin-S, specifically designed to enhance its performance in microscopy image segmentation tasks. These modifications seek to induce much more local context and finer details, which is necessary when dealing with microscopy images.

\subsection{Comparison of Transformer-based Models}

\begin{table}[t]
\centering
\setlength{\tabcolsep}{0.2cm}
\caption{Performance results of UNETR, Swin-UPerNet with Swin-S and Swin-B as the backbones, and Segment Anything Model operating in three different modes (the number of point and bounding box prompts are equal to the number of ground truth instances in each test image) compared to U-Net across datasets. Each row represents the model, while each column represents the obtained F1 and IoU values on each dataset. The best scores are highlighted in \textbf{bold}, and the second best scores are \underline{underlined}.}
\label{tab:basic_model_comparison}
\begin{tabular}{@{}lcccccccc@{}}
\toprule
\multirow{3}{*}{Models} & \multicolumn{2}{c}{\multirow{2}{*}{LIVECell}} & \multicolumn{2}{c}{Seven}      & \multicolumn{2}{c}{\multirow{2}{*}{MoNuSeg}} & \multicolumn{2}{c}{Electron} \\
      & \multicolumn{2}{c}{}                          & \multicolumn{2}{c}{Cell Lines} & \multicolumn{2}{c}{}                         & \multicolumn{2}{c}{Microscopy} \\
 & \textbf{F1} & \textbf{IoU} & \textbf{F1} & \textbf{IoU} & \textbf{F1} & \textbf{IoU} & \textbf{F1} & \textbf{IoU} \\
\midrule
U-Net     & \underline{0.92} & \underline{0.86} & \textbf{0.81} & \textbf{0.70} & 0.80 & 0.68 & 0.92 & \textbf{0.88} \\
UNETR     & \textbf{0.93} & \textbf{0.87} & \underline{0.80} & \underline{0.68} & 0.82 & 0.70 & 0.83 & 0.75 \\
Swin-S          & \underline{0.92} & 0.85 & 0.75 & 0.61 & 0.82 & 0.70 & \textbf{0.93} & \textbf{0.88} \\
Swin-B  & \underline{0.92} & \underline{0.86} & 0.77 & 0.64 & \underline{0.83} & \underline{0.71} & \textbf{0.93} & \textbf{0.88} \\
SAM (Bounding Box)   & 0.86 & 0.76 & 0.78 & 0.64 & \textbf{0.88} & \textbf{0.79} & 0.87 & 0.80 \\
SAM (Point Prompts)  & 0.57 & 0.46 & 0.27 & 0.16 & 0.71 & 0.57 & 0.61 & 0.52 \\
SAM (Automatic Mode) & 0.46 & 0.35 & 0.17 & 0.10 & 0.66 & 0.50 & 0.77 & 0.67 \\
\bottomrule
\end{tabular}
\end{table}

We fine-tuned and evaluated U-Net, UNETR, and Swin-UPerNet on each dataset separately, following our pipeline outlined in Section~\textcolor{red}{\ref{training_pipeline}}. In contrast, we assessed SAM's out-of-the-box performance without fine-tuning to evaluate its immediate usability. We provided the bounding boxes and point prompts equal to the number of instances in each test image for a fair comparison. The results, detailed in Table \textcolor{red}{\ref{tab:basic_model_comparison}}, show that U-Net consistently performs well across all datasets, achieving the highest IoU of 0.88 on the Electron Microscopy dataset, setting a strong baseline for other models. UNETR generally matches the performance of U-Net but lags on the Electron Microscopy dataset with the 0.75 IoU score. Both small and basic versions of Swin-UPerNet are behind U-Net and UNETR across almost all of the datasets, except for the Electron Microscopy, showing the same results as the U-Net. These observations highlight that the traditional CNN approach remains good and robust. Segment Anything Model, when using bounding box prompts, shows reasonable performance but does not achieve the same levels as fine-tuned models. Although bounding box prompts provide decent test scores, performance decreases when switching to point prompts or using automatic mode, particularly on the Seven Cell Lines dataset.

\subsection{Comparison of Swin-UPerNet Modifications}

\begin{table}[t]
    \centering
    \caption{Performance results of Swin-UPerNet (Swin-S) modifications compared to U-Net and original Swin-UPerNet (Swin-S and Swin-B) across datasets. Each row represents the model or the modification, while each column represents the obtained F1 and IoU values on each dataset. The best scores are highlighted in \textbf{bold}, and the second best scores are \underline{underlined}.}
    \label{swin_mods_res}
    \setlength{\tabcolsep}{0.20cm}
    \begin{tabular}{lcccccccc}
    \toprule
        \textbf{Models} & \multicolumn{2}{c}{\multirow{2}{*}{LIVECell}} & \multicolumn{2}{c}{Seven}      & \multicolumn{2}{c}{\multirow{2}{*}{MoNuSeg}} & \multicolumn{2}{c}{Electron} \\
      & \multicolumn{2}{c}{}                          & \multicolumn{2}{c}{Cell Lines} & \multicolumn{2}{c}{}                         & \multicolumn{2}{c}{Microscopy} \\
 & \textbf{F1} & \textbf{IoU} & \textbf{F1} & \textbf{IoU} & \textbf{F1} & \textbf{IoU} & \textbf{F1} & \textbf{IoU} \\ 
 \midrule
        U-Net  & 0.92 & 0.86 & 0.81 & 0.70 & 0.80 & 0.68 & 0.92 & 0.88 \\
        Swin-S & 0.92 & 0.85 & 0.75 & 0.61 & \underline{0.82} & 0.70 & 0.93 & 0.88 \\
        Swin-B  & 0.92 & 0.86 & 0.77 & 0.64 & \textbf{0.83} & \textbf{0.71} & 0.93 & 0.88 \\
        \midrule
        Swin-S-PS2     &  \textbf{0.93} & \underline{0.87} & 0.81 & 0.69 & \underline{0.82} & 0.70 & 0.94 & 0.89 \\
        Swin-S-Conv    & 0.92 & 0.86 & 0.77 & 0.64 & 0.81 & 0.69 & \textbf{0.95} & \underline{0.90} \\
        Swin-S-TB   & \textbf{0.93} & \underline{0.87} & \underline{0.83} & \underline{0.72} & \underline{0.82} & 0.70 & 0.91 & 0.86 \\
        Swin-S-TB-Skip      & \textbf{0.93} & \textbf{0.88} & \textbf{0.84} & \textbf{0.74} & \underline{0.82} & \textbf{0.71} & \textbf{0.95} & \textbf{0.91} \\
        Swin-S-Pyramid & 0.91 & 0.85 & 0.80 & 0.67 & 0.80 & 0.67 & 0.90 & 0.84 \\
        \bottomrule
    \end{tabular}
\end{table}

We fine-tuned all of the designed modifications on each dataset separately, utilising the proposed train pipeline, described in Section~\textcolor{red}{\ref{training_pipeline}} and draw a comparison between the original U-Net and Swin-UPerNet proposed architectures, aiming to increase its performance in microscopy image segmentation. All of the presented modifications were based on the Swin-S architecture. From Table \textcolor{red}{\ref{swin_mods_res}}, we can see that our Swin-S-TB-Skip modification excels across almost all datasets, surpassing U-Net and the original Swin-UPerNet models, both small (Swin-S) and basic (Swin-B) versions, achieving higher IoU score. Apart from this, we can see a notable increase in the performance of Swin-S-TB-Skip compared to Swin-UPerNet on the Seven Cell Lines dataset, which contains brightfield images with the nuclei as a target. We consider this modification to be our best among the others. 

\subsection{FLOPs and Parameters of the Models}

\begin{table}[t]
    \centering
    \caption{Model Parameters and FLOPs. We calculated the number of FLOPs by passing the 3-channel image of size $224\times224$ to the model. Swin-S-PS2 denotes our modification of Swin-UPerNet with decreased patch size to $2\times2$. Swin-S-TB-Skip denotes our best modification, with the extension of the encoder, decrease in patch size, replacement of interpolation and addition of skip connection. We cannot provide FLOPs for SAM as it depends on the amount of prompts provided to the model, which can greatly vary.}
    \setlength{\tabcolsep}{0.2cm}
    \label{tab:params_flops}
    \begin{tabular}{lcc}
        \toprule
        \textbf{Model} & \textbf{Params (M)} & \textbf{FLOPs (G)} \\
        \midrule
        U-Net & 24.4 & 12 \\
        UNETR & 111.7 & 234 \\
        Swin-B & 121.1 & 128 \\
        Swin-S & 81.1 & 98 \\
        Swin-S-PS2 (ours)& 81.1 & 390 \\
        Swin-S-TB-Skip (ours) & 82.1 & 452 \\
        SAM & 93.7 & --- \\
        \bottomrule
    \end{tabular}
\end{table}

Here, we provide an overview of the FLOPs and parameters for UNETR, U-Net, SAM, Swin-UPerNet, and its best modification --- Swin-S-TB-Skip. The FLOPs were calculated using a 3-channel image of $224\times224$ as the input to each model. Table \textcolor{red}{\ref{tab:params_flops}} shows that U-Net has the least amount of FLOPs among all other models.

Another notable observation is that Swin-S-TB has almost 4.5 times more FLOPs than the original Swin-UPerNet. While this may sound alarming, the reason behind it is simple. As the patch size decreases to $2\times2$, the number of patches in the image increases, leading to a fourfold increase in the size of the attention matrix. Although this modification requires many more FLOPs to run, it still fits within the memory constraints of the same GPU. We could not provide FLOPs for the SAM model, as it depends on the number of prompts passed to the model by the user, which can greatly vary.

\section{Discussion}
Our experimental results offer an overview of the capabilities of modern transformer-based models --- UNETR, Swin-UPerNet, and SAM in recognizing and segmenting various objects and structures within microscopy images across different modalities. We compared these models to the established U-Net model and evaluated their performance. Figure \textcolor{red}{\ref{fig:examples}} provides examples of the predictions of our best modification - Swin-TB-Skip compared to the UNETR, U-Net, and Segment Anything Model across all of the datasets. The results in Table \textcolor{red}{\ref{tab:basic_model_comparison}} show that U-Net remains a strong contender for semantic segmentation in microscopy images. While UNETR and Swin-UPerNet generally match U-Net's performance, their significant computational demands make them less practical for real-world applications. 

The original Swin-UPerNet slightly falls behind the U-Net across most of the datasets, but several innovative modifications we introduced greatly enhanced its performance. Modifications such as extending the encoder, replacing interpolation layers in the decoder, adding an extra skip connection, and reducing patch size aim to improve local context modelling. This is crucial for better segmentation quality in microscopy images, where object and structure variability is high. Our top-performing modification, Swin-S-TB-Skip, showed notable improvements across all datasets. It is also notable, that our best modification surpassed the basic version of the original Swin-UperNet, which has more parameters, highlighting the value and the relevance of our architectural improvements. We emphasize the substantial performance boost on the Seven Cell Lines dataset compared to the original Swin-UPerNet. The brightfield modality and the challenge of segmenting cell nuclei make this dataset especially difficult.

Segment Anything Model, the first foundational segmentation model, has delivered mixed results across various datasets. The model heavily depends on user-defined prompts to achieve improvements in performance, particularly noticeable when bounding boxes are employed in contrast to its baseline automatic segmentation capabilities. This reliance on user input for optimal performance significantly diminishes its utility compared to other models. Without user interaction, SAM’s segmentations are often suboptimal, limiting its direct comparability and competitiveness with automated models that do not require such inputs. Moreover, SAM lacks class awareness, indiscriminately segmenting all detectable objects and structures. This restricts its applicability for specialized tasks, such as cell segmentation, where targeted recognition of specific classes is crucial. Nonetheless, with further developments, such as user interface, SAM could evolve into a valuable tool for interactive annotation.

These findings demonstrate that there is still potential for advancement in transformer-based models. Their unique attention mechanisms hold promise for achieving cutting-edge performance in segmentation tasks. By continuously refining and improving these architectures, we can unlock their full potential and establish new benchmarks in the field.

\section{Conclusion}

In this study, we make two major contributions to the field of microscopy image segmentation. Our first contribution is a detailed comparison between the well-established and popular U-Net model and several innovative transformer-based deep learning models. These include Swin-UPerNet, which features a unique windowed attention mechanism, the Segment Anything Model with its interactive prompt segmentation approach, and UNETR, which blends a traditional U-Net-like decoder with a modern vision transformer encoder. Our evaluations reveal that while these modern transformer-based models perform comparably to U-Net, there is still room for improvement.

Our second major contribution focuses on enhancing the performance of the Swin-UPerNet model. We conducted a series of experiments aimed at increasing its robustness and performance across various microscopy images. The modifications we implemented greatly improved the performance of the model. Our revised version, Swin-S-TB-Skip, on average, outperformed the original Swin-UPerNet and U-Net across all tested microscopy datasets, achieving a higher IoU score.

However, these performance gains come with increased computational demand (see Table \textcolor{red}{\ref{tab:params_flops}}). Future research should, therefore, concentrate on optimising these architectural enhancements for practical applications and integrating them into diverse microscopy image analysis workflows and tools.

\section*{Acknowledgments}
This work was supported by Revvity~\footnote{\url{https://www.revvity.com/}}. Computational resources were provided by the High-Performance Computing Cluster of the University of Tartu~\cite{hpc}. A big thank you to all the members of the Biomedical Computer Vision Lab at the University of Tartu for their continuous support.

\bibliographystyle{splncs04}
\bibliography{main}

\begin{thebibliography}{10}
\providecommand{\url}[1]{\texttt{#1}}
\providecommand{\urlprefix}{URL }
\providecommand{\doi}[1]{https://doi.org/#1}

\bibitem{CS-Unet}
Alrfou, K., Zhao, T., Kordijazi, A.: Cs-unet: A generalizable and flexible segmentation algorithm (Apr 2024). \doi{10.1007/s11042-024-19242-4}, \url{http://dx.doi.org/10.1007/s11042-024-19242-4}

\bibitem{wandb}
Biewald, L.: Experiment tracking with weights and biases (2020), \url{https://www.wandb.com/}, software available from wandb.com

\bibitem{albumnetations}
Buslaev, A., Iglovikov, V.I., Khvedchenya, E., Parinov, A., Druzhinin, M., Kalinin, A.A.: Albumentations: Fast and flexible image augmentations. Information  \textbf{11}(2) (2020). \doi{10.3390/info11020125}, \url{https://www.mdpi.com/2078-2489/11/2/125}

\bibitem{Swin-Unet}
Cao, H., Wang, Y., Chen, J., Jiang, D., Zhang, X., Tian, Q., Wang, M.: Swin-unet: Unet-like pure transformer for medical image segmentation. In: Computer Vision – ECCV 2022 Workshops: Tel Aviv, Israel, October 23–27, 2022, Proceedings, Part III. p. 205–218. Springer-Verlag, Berlin, Heidelberg (2023). \doi{10.1007/978-3-031-25066-8_9}, \url{https://doi.org/10.1007/978-3-031-25066-8_9}

\bibitem{transunet}
Chen, J., Lu, Y., Yu, Q., Luo, X., Adeli, E., Wang, Y., Lu, L., Yuille, A.L., Zhou, Y.: Transunet: Transformers make strong encoders for medical image segmentation. arXiv preprint arXiv:2102.04306  (2021)

\bibitem{hybrid_trans_unet}
Chi, J., Li, Z., Sun, Z., Yu, X., Wang, H.: Hybrid transformer unet for thyroid segmentation from ultrasound scans. Computers in Biology and Medicine  \textbf{153},  106453 (2023). \doi{https://doi.org/10.1016/j.compbiomed.2022.106453}, \url{https://www.sciencedirect.com/science/article/pii/S0010482522011611}

\bibitem{ImageNet}
Deng, J., Dong, W., Socher, R., Li, L.J., Li, K., Fei-Fei, L.: Imagenet: A large-scale hierarchical image database. In: 2009 IEEE Conference on Computer Vision and Pattern Recognition. pp. 248--255 (2009). \doi{10.1109/CVPR.2009.5206848}

\bibitem{ViT}
Dosovitskiy, A., Beyer, L., Kolesnikov, A., Weissenborn, D., Zhai, X., Unterthiner, T., Dehghani, M., Minderer, M., Heigold, G., Gelly, S., Uszkoreit, J., Houlsby, N.: An image is worth 16x16 words: Transformers for image recognition at scale. In: International Conference on Learning Representations (2021), \url{https://openreview.net/forum?id=YicbFdNTTy}

\bibitem{livecell}
Edlund, C., Jackson, T.R., Khalid, N., Bevan, N., Dale, T., Dengel, A., Ahmed, S., Trygg, J., Sjögren, R.: Livecell—a large-scale dataset for label-free live cell segmentation (Aug 2021). \doi{10.1038/s41592-021-01249-6}, \url{http://dx.doi.org/10.1038/s41592-021-01249-6}

\bibitem{sunet}
Fan, C.M., Liu, T.J., Liu, K.H.: Sunet: swin transformer unet for image denoising. In: 2022 IEEE International Symposium on Circuits and Systems (ISCAS). pp. 2333--2337. IEEE (2022)

\bibitem{microscopy_intro}
Fishman, D., Salumaa, S., Majoral, D., Laasfeld, T., Peel, S., Wildenhain, J., Schreiner, A., Palo, K., Parts, L.: Practical segmentation of nuclei in brightfield cell images with neural networks trained on fluorescently labelled samples (Jun 2021). \doi{10.1111/jmi.13038}, \url{http://dx.doi.org/10.1111/jmi.13038}

\bibitem{swinUNETR}
Hatamizadeh, A., Nath, V., Tang, Y., Yang, D., Roth, H.R., Xu, D.: Swin unetr: Swin transformers for semantic segmentation of brain tumors in mri images. In: Crimi, A., Bakas, S. (eds.) Brainlesion: Glioma, Multiple Sclerosis, Stroke and Traumatic Brain Injuries. pp. 272--284. Springer International Publishing, Cham (2022)

\bibitem{UNETR}
Hatamizadeh, A., Tang, Y., Nath, V., Yang, D., Myronenko, A., Landman, B., Roth, H.R., Xu, D.: Unetr: Transformers for 3d medical image segmentation. In: 2022 IEEE/CVF Winter Conference on Applications of Computer Vision (WACV). pp. 1748--1758 (2022). \doi{10.1109/WACV51458.2022.00181}

\bibitem{PPM}
He, K., Zhang, X., Ren, S., Sun, J.: Spatial pyramid pooling in deep convolutional networks for visual recognition. In: Fleet, D., Pajdla, T., Schiele, B., Tuytelaars, T. (eds.) Computer Vision -- ECCV 2014. pp. 346--361. Springer International Publishing, Cham (2014)

\bibitem{smp}
Iakubovskii, P.: Segmentation models pytorch (2019), \url{https://github.com/qubvel/segmentation_models.pytorch}

\bibitem{opencv}
Itseez: Open source computer vision library. \url{https://github.com/itseez/opencv} (2015)

\bibitem{sam}
Kirillov, A., Mintun, E., Ravi, N., Mao, H., Rolland, C., Gustafson, L., Xiao, T., Whitehead, S., Berg, A.C., Lo, W.Y., Dollár, P., Girshick, R.: Segment anything. In: 2023 IEEE/CVF International Conference on Computer Vision (ICCV). pp. 3992--4003 (2023). \doi{10.1109/ICCV51070.2023.00371}

\bibitem{monuseg_1}
Kumar, N., Verma, R., Anand, D., Zhou, Y., Onder, O.F., Tsougenis, E., Chen, H., Heng, P.A., Li, J., Hu, Z., Wang, Y., Koohbanani, N.A., Jahanifar, M., Tajeddin, N.Z., Gooya, A., Rajpoot, N., Ren, X., Zhou, S., Wang, Q., Shen, D., Yang, C.K., Weng, C.H., Yu, W.H., Yeh, C.Y., Yang, S., Xu, S., Yeung, P.H., Sun, P., Mahbod, A., Schaefer, G., Ellinger, I., Ecker, R., Smedby, O., Wang, C., Chidester, B., Ton, T.V., Tran, M.T., Ma, J., Do, M.N., Graham, S., Vu, Q.D., Kwak, J.T., Gunda, A., Chunduri, R., Hu, C., Zhou, X., Lotfi, D., Safdari, R., Kascenas, A., O’Neil, A., Eschweiler, D., Stegmaier, J., Cui, Y., Yin, B., Chen, K., Tian, X., Gruening, P., Barth, E., Arbel, E., Remer, I., Ben-Dor, A., Sirazitdinova, E., Kohl, M., Braunewell, S., Li, Y., Xie, X., Shen, L., Ma, J., Baksi, K.D., Khan, M.A., Choo, J., Colomer, A., Naranjo, V., Pei, L., Iftekharuddin, K.M., Roy, K., Bhattacharjee, D., Pedraza, A., Bueno, M.G., Devanathan, S., Radhakrishnan, S., Koduganty, P., Wu, Z., Cai, G., Liu, X., Wang, Y., Sethi, A.:
  A multi-organ nucleus segmentation challenge. IEEE Transactions on Medical Imaging  \textbf{39}(5),  1380--1391 (2020). \doi{10.1109/TMI.2019.2947628}

\bibitem{monuseg_2}
Kumar, N., Verma, R., Sharma, S., Bhargava, S., Vahadane, A., Sethi, A.: A dataset and a technique for generalized nuclear segmentation for computational pathology. IEEE Transactions on Medical Imaging  \textbf{36}(7),  1550--1560 (2017). \doi{10.1109/TMI.2017.2677499}

\bibitem{FPN}
Lin, T.Y., Dollár, P., Girshick, R., He, K., Hariharan, B., Belongie, S.: Feature pyramid networks for object detection. In: 2017 IEEE Conference on Computer Vision and Pattern Recognition (CVPR). pp. 936--944 (2017). \doi{10.1109/CVPR.2017.106}

\bibitem{FocalLoss}
Lin, T.Y., Goyal, P., Girshick, R., He, K., Dollár, P.: Focal loss for dense object detection  (2017). \doi{10.48550/ARXIV.1708.02002}, \url{https://arxiv.org/abs/1708.02002}

\bibitem{SWIN}
Liu, Z., Lin, Y., Cao, Y., Hu, H., Wei, Y., Zhang, Z., Lin, S., Guo, B.: Swin transformer: Hierarchical vision transformer using shifted windows. In: 2021 IEEE/CVF International Conference on Computer Vision (ICCV). pp. 9992--10002 (2021). \doi{10.1109/ICCV48922.2021.00986}

\bibitem{pytroch}
Paszke, A., Gross, S., Massa, F., Lerer, A., Bradbury, J., Chanan, G., Killeen, T., Lin, Z., Gimelshein, N., Antiga, L., Desmaison, A., Kopf, A., Yang, E., DeVito, Z., Raison, M., Tejani, A., Chilamkurthy, S., Steiner, B., Fang, L., Bai, J., Chintala, S.: Pytorch: An imperative style, high-performance deep learning library. In: Advances in Neural Information Processing Systems 32, pp. 8024--8035. Curran Associates, Inc. (2019), \url{http://papers.neurips.cc/paper/9015-pytorch-an-imperative-style-high-performance-deep-learning-library.pdf}

\bibitem{Unet}
Ronneberger, O., Fischer, P., Brox, T.: U-net: Convolutional networks for biomedical image segmentation. In: Navab, N., Hornegger, J., Wells, W.M., Frangi, A.F. (eds.) Medical Image Computing and Computer-Assisted Intervention -- MICCAI 2015. pp. 234--241. Springer International Publishing, Cham (2015)

\bibitem{SSFormer}
Shi, W., Xu, J., Gao, P.: Ssformer: A lightweight transformer for semantic segmentation. In: 2022 IEEE 24th International Workshop on Multimedia Signal Processing (MMSP). pp.~1--5 (2022). \doi{10.1109/MMSP55362.2022.9949177}

\bibitem{DiceLoss}
Sudre, C.H., Li, W., Vercauteren, T., Ourselin, S., Jorge~Cardoso, M.: Generalised dice overlap as a deep learning loss function for highly unbalanced segmentations. In: Cardoso, M.J., Arbel, T., Carneiro, G., Syeda-Mahmood, T., Tavares, J.M.R., Moradi, M., Bradley, A., Greenspan, H., Papa, J.P., Madabhushi, A., Nascimento, J.C., Cardoso, J.S., Belagiannis, V., Lu, Z. (eds.) Deep Learning in Medical Image Analysis and Multimodal Learning for Clinical Decision Support. pp. 240--248. Springer International Publishing, Cham (2017)

\bibitem{hpc}
{University of Tartu}: High performance computing center, institute of computer science (2018). \doi{10.23673/PH6N-0144}

\bibitem{attention_is_all_you_need}
Vaswani, A., Shazeer, N., Parmar, N., Uszkoreit, J., Jones, L., Gomez, A.N., Kaiser, L.u., Polosukhin, I.: Attention is all you need. In: Guyon, I., Luxburg, U.V., Bengio, S., Wallach, H., Fergus, R., Vishwanathan, S., Garnett, R. (eds.) Advances in Neural Information Processing Systems. vol.~30. Curran Associates, Inc. (2017), \url{https://proceedings.neurips.cc/paper_files/paper/2017/file/3f5ee243547dee91fbd053c1c4a845aa-Paper.pdf}

\bibitem{Hydra}
Yadan, O.: Hydra - a framework for elegantly configuring complex applications. Github (2019), \url{https://github.com/facebookresearch/hydra}

\bibitem{em}
Yildirim, B., Cole, J.M.: Bayesian particle instance segmentation for electron microscopy image quantification (Mar 2021). \doi{10.1021/acs.jcim.0c01455}, \url{http://dx.doi.org/10.1021/acs.jcim.0c01455}

\bibitem{SETR}
Zheng, S., Lu, J., Zhao, H., Zhu, X., Luo, Z., Wang, Y., Fu, Y., Feng, J., Xiang, T., Torr, P.H., Zhang, L.: Rethinking semantic segmentation from a sequence-to-sequence perspective with transformers. In: CVPR (2021)

\end{thebibliography}
\end{document}